\documentstyle[amsfonts,aps]{revtex}
\newcommand{\hodge}[1]{\,*#1}
\newcommand{\lie}{{\cal L}}
\begin{document}
\title{Analyzing Charges in even dimensions}
\author{Rodrigo Aros}
\address{Centro de Estudios Cient\'{\i }ficos (CECS), Casilla 1469,
Valdivia, Chile.\\
Universidad Nacional Andres Bello, Sazie 2320, Santiago,Chile }
\date{\today}
\maketitle
\begin{abstract}
Lanczos-Lovelock theories of gravity, in its first order version, are studied on asymptotically
locally anti de Sitter spaces. It is shown that thermodynamics satisfies the standard behavior and
an expression for entropy is found for this formalism. Finally a short analysis of the algebra of
conserved charges is displayed.
\end{abstract}

\section{Introduction}
Standard general relativity is given by the Einstein Hilbert action (plus a cosmological constant).
This theory has as dynamical field the metric and by definition is torsionless, which yields to
second order field equations. In fact, in three and four dimensions including higher power of the
Riemann tensor unavoidably yields to higher order field equations\footnote{In four dimensions there
is a second order polynomial of the Riemann tensor, the four dimensional Euler density, which is a
topological invariant thus it does not contribute to the field equations.}. However in higher
dimensions to have second order field equations is not restricted to the Einstein theory, but there
are others sensible Lagrangians which contain higher power of the Riemann tensor and yet have
second order field equations. This fact was first noted by Lanczos\cite{Lanczos:1975su} in five
dimensions and lately generalized by Lovelock to higher dimensions\cite{Lovelock:1971yv}.

Nonetheless, a major drawback of the metric formulation is that spinors can not be incorporated in
this formalism, thus another approach has to be taken. In this way, the first order formalism shows
up. On a manifold ${\cal{M}}$, the fields are the orthonormal basis of the cotangent space, $e^a$
(also called vielbein), and the -spin- connection $w^{ab}$, which defines the Lorentz covariant
derivatives. Since $e^a$ and $w^{ab}$ are $1$-forms that transform under local Lorentz group, the
Lagrangian to be constructed is manifestly invariant under diffeomorphisms and additionally should
be a scalar under Lorentz transformations. This kind of theories has been vastly studied in the
literature, for a review see Ref.\cite{Troncoso:1999pk} (and references therein). Finally,
recalling that in this formalism $e^a$ and $w^{ab}$ are independent fields, this is a genuine first
order theory fields\cite{Muller-Hoissen:1991ht}, therefore there are issues to be considered. In
particular to define boundary conditions that do not over determine the fields and simultaneously
that define a proper action principle is a hard task. For even dimensions it was introduced in
\cite{Aros:1999id}, and thereafter extended in \cite{Aros:1999kt}, a new boundary condition, the
asymptotically locally AdS condition, which satisfies both criteria. This issue will be discussed
in detail at section \ref{Bconditions}. From now on these Lagrangians with higher power of the
Riemann tensor, written in first order formalism, will be named \textbf{LL} Lagrangians, after
Lanczos and Lovelock.

Recently the Noether method was successfully applied to LL Lagrangians defined on asymptotically
locally AdS (\textbf{ALAdS}) spaces \cite{Aros:1999id,Aros:1999kt}, and unlike the Komar's
prescription, these Noether charges indeed reproduce the Hamiltonian charges. This remarkable
feature leads to attempt a deeper analysis of the LL theories, unfortunately any LL Lagrangian has
in principle a singular symplectic matrix, which in turn makes a general Hamiltonian approach (See
Ref.\cite{Regge:1974zd}) hard to achieve, and as far unknown. One option is to follow the
\emph{covariant phase space method}. Formally this method reproduces from a purely Lagrangian
framework the Hamiltonian formalism. The introduction of this method to gravitation was made in
\cite{Lee:1990nz}, where it was applied to the metric formulation of EH gravity reproducing the
well known Hamiltonian results. Also, in \cite{Carlip:1999cy}, the analysis in $2+1$ dimensions of
the algebra of conserved charges at the horizon was performed by using this method. Finally it is
worth noting that the thermodynamical analysis of the LL Lagrangians in the metric formalism, which
was first established in \cite{Whitt:1988ax} using Gibbons-Hawking regularization procedure, was
also analyzed in \cite{Jacobson:1993xs} using this approach.

In this work the covariant phase space method is applied to the LL theories in their first order
versions. In the first part of this work the method and some general ideas of the first order
gravitational theories are described, afterward, it is shown how the first law of the black hole
thermodynamics holds for first order gravitational theories on asymptotically locally AdS
spaces\cite{Whitt:1988ax,Jacobson:1993xs}. Thereafter it is displayed as an example the explicit
calculation for the Born-Infeld Lagrangian\cite{Banados:1994ur}. Finally a first approach to the
analysis of conserved charges algebras is displayed.

In this work an even ($2n$) dimensional manifold $\mathcal{M}$ is considered. For simplicity solely
black hole geometries are studied, and following the Hamiltonian standard case $\mathcal{M}$ has
topology of ${\mathbb{R}}\times \Sigma$, where $\Sigma$ is a space like sub manifold. In addition,
to characterize the geometry, $\partial\Sigma_\infty$ and $\partial\Sigma_H$ will denote the
spatial infinity and event horizon respectively, thus the boundary of ${\mathcal{M}}$ is
\[
\partial{\mathcal{M}}=\partial \Sigma \times {\mathbb{R}}=\partial \Sigma_H \times
{\mathbb{R}}\cup\partial \Sigma_\infty \times {\mathbb{R}}.
\]
The differential forms language will be assumed, so exterior product ($\wedge$) between
differential forms will be understood. In particular any field in this work is regarded as a
$1$-form, as well as the Lagrangian as a $2n$-form. Finally the name vector will denote only
geometrical objects as $\xi=\xi^{\mu} \partial_{\mu}$.

\section{Presymplectic structure for an arbitrary theory}\label{SSAT}

Any infinitesimal transformation of a field $\phi(x)$ can be split as a local functional
transformation plus a diffeomorphism, given by $x' = x + \xi(x)$ (See \cite{Ramond:1989yd}). This
can be done as follows,
\begin{equation}\label{SplitingTheTransf}
\delta \phi(x)= \phi'(x') - \phi(x)= \phi'(x') - \phi'(x) + \phi'(x) - \phi(x)
\end{equation}
where $\phi'(x) - \phi(x)= \delta_{0} \phi(x)$ is a local functional transformation, and $\phi'(x')
- \phi'(x) = {\cal L}_\xi \phi$. Finally one obtains that $\delta\phi = \delta_{0}\phi + {\cal
L}_\xi \phi$. Applying those definitions to any Lagrangian theory one obtains that\footnote{The
operator $I_\xi$ acts over a differential form, $A_{p}={1\over (p!)} A_{i_{1}\ldots
i_{p}}dx^{i_{1}}\wedge\ldots\wedge dx^{i_{p}}$, as
\begin{equation}
  I_{\xi} A_{p} = \xi^{j}{1\over (p-1)!}A_{j i_{2}\ldots i_{p}}dx^{i_{2}}\wedge\ldots\wedge dx^{i_{p}}
\end{equation}

In term of this operator the Lie derivative acting over a differential form can be written as
${\cal L}_\xi A_{p} = (dI_\xi + I_\xi d) A_{p}$}
\begin{equation}\label{LagrangianVariation}
  \delta {\mathbf{L}} = (E.M.)\delta_{0} \phi + d \Theta(\phi,\delta_{0}\phi) + d I_\xi
  {\mathbf{L}},
\end{equation}
where $(E.M.)$ represents the \textit{equations of motion}.

A symmetry is defined as a change in the field configuration that does not alter the field
equations. To satisfy that $\delta\phi$ has to be such that the Lagrangian changes in a total
derivative, i.e., $\delta {\mathbf{L}} = d \alpha$. For the sake of clarity when a transformation
$\delta$ corresponds to a symmetry, $\delta_{0}$ will be denoted $\hat{\delta}$. In this case the
Noether method states that, evaluated on the solution, the current
\begin{equation}\label{currentdensity}
 \hodge{{\bf J}_\xi} = \Theta(\hat{\delta} \phi,\phi) + I_\xi{\bf L}-\alpha,
\end{equation}
satisfies $d(\hodge{{\bf J}_\xi})=0$.

On the other hand, given that differential forms are manifestly invariant under diffeomorphisms,
i.e., $\delta \phi = 0$, the associated conserved current is obtained by substituting in
Eq.(\ref{currentdensity}), $\hat{\delta} \phi = - {\cal L}_\xi \phi$ and $\alpha=0$. Therefore the
Noether current for diffeomorphism is
\begin{equation}\label{Diffcurrentdensity}
\hodge{{\bf J}_\xi} = \Theta(-\lie_{\xi} \phi,\phi) + I_\xi{\bf L}.
\end{equation}
By using this current one can define a conserved charge, however in principle it will be only of
relevance, i.e. a physical charge,  if $\xi$ corresponds to a diffeomorphism that generates an
isometry of the solution, namely a Killing vector.

The presymplectic form $\Xi$ is defined by \cite{Lee:1990nz,Carlip:1999cy}
\begin{equation}\label{Symplecticmatrix}
  \Xi(\phi,\delta_1\phi,\delta_2\phi) = \int_\Sigma \delta_1 \Theta(\phi,\delta_2\phi)-\delta_2
  \Theta(\phi,\delta_1\phi),
\end{equation}
where \textbf{$\delta_1$ and $\delta_2$ correspond to functional variations of the fields, i.e.,
they correspond to independents $\delta_{0}$'s.}

Inside the space of configurations $\mathcal{F}$ exists the subspace of classical solutions
$\bar{\mathcal{F}}$. In this subspace one has that $\delta_{1} \hodge{{\bf J}_\xi} = \delta_{1}
\Theta(-{\cal L}_\xi \phi, \phi) + I_\xi d\Theta(\delta_{1}\phi,\phi)$ which, by using $I_\xi d=
\lie_\xi - dI_\xi$, implies that
\begin{equation}\label{ExtensionofCharge}
\left( \delta_{1} \Theta(-{\cal L}_\xi \phi,\phi)+ \lie_\xi \Theta(\delta_{1}\phi,\phi)\right) =
\delta_{1} \hodge{{\bf J}_\xi} + dI_\xi \Theta(\delta_{1}\phi,\phi).
\end{equation}
The left hand side, once it is integrated on $\Sigma$, can be identified as the presymplectic form
evaluated for the variations $(\delta_{1}\phi,-\lie_\xi\phi)$. Simultaneously this presymplectic
form (Eq.(\ref{ExtensionofCharge})) can be interpreted as the Hamiltonian variation $\delta_{1}$ of
the generator of diffeomorphisms\footnote{Recalling that expressions are on-shell the generator of
diffeomorphisms, $G(\xi)=H_{\mu}\xi^{\mu}+ C(\xi)$, is simply given by $G(\xi)=C(\xi)$}
$G(\xi)$\cite{Lee:1990nz}. Therefore,
\begin{equation}\label{VariationofTheCharge}
  \Xi(\phi,\delta_{1}\phi,-\lie_\xi\phi)= \delta_{1} G(\xi)= \int_\Sigma \delta_{1} \hodge{{\bf J}_\xi} + dI_\xi
\Theta(\delta_{1}\phi,\phi).
\end{equation}

\section{First Order Gravity}
\subsection{General ideas}\label{Kgravities}
Recalling that $\mathcal{M}$ is $2n$-dimensional manifold, the first order LL Lagrangian is
constructed in terms of Riemann curvatures and vielbein as
\begin{equation}\label{initialaction}
{\mathbf{L}} = \kappa\sum_{p=0}^{n-1} \alpha_p \,\epsilon_{a_1\ldots a_{2n}} \left[(R)^{p}
\left(\frac{e}{l}\right)^{2n-2p}\right]^{a_1 \ldots a_{2n}},
\end{equation}
where
\[\left[(R)^{p} \left(\frac{e}{l}\right)^{2n-2p}\right]^{a_1 \ldots a_{2n}} = R^{a_1 a_2}\ldots
R^{a_{2p-1} a_{2p}} \frac{e}{l}^{a_{2p+1}} \ldots \frac{e}{l}^{a_{2n}},
\]
$R^{ab}=dw^{ab} + w^a_{\,\,c}w^{cb}$ is the Riemann two-form of curvature, $\alpha_p$ arbitrary
constants and $\kappa$ a normalization constant\footnote{ In particular the four dimensional
Einstein Hilbert action with a negative cosmological constant is given by
\begin{equation}\label{EH1}
  S_{EH} = \frac{1}{32\pi G} \int_{\mathcal{M}} R^{ab} e^c e^d \epsilon_{abcd}
   + \frac{1}{2l^2}e^a e^b e^c e^d \epsilon_{abcd},
\end{equation}
where $l^2 = -3\Lambda$, being $\Lambda$ the cosmological constant.}.  A theory of gravity, as the
one described by Eq.(\ref{initialaction}), has multiple background configurations, since this
Lagrangian does not have a single cosmological constant. This problem was analyzed in
Ref.\cite{Crisostomo:2000bb}, where the arbitrary coefficients $\alpha_{p}$ were restricted such
that having a single negative cosmological constant, in this case $l$ is related with the value of
that cosmological constant. From now on, it will be considered this kind of theories only, however
those set of $\{\alpha_{p}\}$ will be only implicitly assumed during discussion, since besides that
they constrain the LL Lagrangians to have a single cosmological constant no significant
simplification is obtained by substituting them.

The variation of the Lagrangian is given by,
\begin{equation}\label{Variation}
\delta_{0}{\mathbf{L}} = G_f\delta_{0} e^f + D(\delta_{0} w^{ab}) \tau_{ab}
\end{equation}
where
\begin{equation}\label{GraviationalEOM}
G_f=\frac{\overleftarrow{\partial}{\mathbf{L}}}{\partial e^f} = \kappa\sum_{p=0}^{n-1}
(2n-2p)\alpha_p \,\epsilon_{a_1\ldots a_{2n-1}f} \left[(R)^{p}
\left(\frac{e}{l}\right)^{2n-2p-1}\right]^{a_1 \ldots a_{2n-1}}
\end{equation}
is a first field equation and
\begin{equation}
\tau_{ab}=\frac{\partial{\mathbf{L}}}{\partial R^{ab}} = \kappa\sum_{p=0}^{n-1} p \alpha_p
\,\epsilon_{ab a_3\ldots a_{2n}} \left[(R)^{p-1} \left(\frac{e}{l}\right)^{2n-2p}\right]^{a_3
\ldots a_{2n}}\label{firstvariation}.
\end{equation}

Using the Stoke theorem the second term in Eq.(\ref{Variation}) can be split such that obtaining
the second field equation $D(\tau_{ab})=0 $ and the boundary term
\begin{equation}\label{GraviationalTheta}
\Theta(\delta_{0} w^{ab} w^{ab} e^d)=\kappa\sum_{p=0}^{n-1} p \alpha_p \,\epsilon_{a_1\ldots
a_{2n}} \left[ \delta_{0} w (R)^{p-1} \left(\frac{e}{l}\right)^{2n-2p}\right]^{a_1 \ldots a_{2n}}.
\end{equation}
In even dimensions, the solution for $D(\tau_{ab})=0$ with minimal requirements is given by a
torsion tensor $T^{a}=de^{a}+ w^{a}_{\,\,\,b}e^{b}=0$. It must be stressed, however, that in odd
dimensions there is a particular set of $\{\alpha_{p}\}$ that gives rise to Chern-Simons action,
for which $T^{a}\neq 0$ is not an over restricted condition, and actually a torsionless solution is
not the most general case that solves $D(\tau_{ab})=0$ (See \cite{Troncoso:1999pk}). Even though
under $T^{a}=0$ the solutions are indeed solutions of the metric formalism, it must be emphasized
that this is a dynamical result and does not mean that both theories are equivalent.

The Noether current, associated to the invariance under diffeomorphisms, can be computed by
substituting Eq.(\ref{GraviationalTheta}) into Eq.(\ref{Diffcurrentdensity}). Therefore,
\begin{equation}\label{GravitationalCharge}
   \hodge{{\bf J}_\xi} = -d\left(I_\xi w^{ab} \tau_{ab}\right).
\end{equation}
That $\hodge{{\bf J}}$ is a closed form directly implies that $d(\hodge{{\bf J}_\xi})=0$ is
satisfied for any $\xi$ and the conserved charges depend only on boundary information. However it
is worth noting that in order to define a physical conserved charge from $\hodge{{\bf J}_\xi}$ is
necessary that $\xi$ to be an isometry of the solution, namely a Killing vector. From now on $\xi$
will be considered a Killing vector.

\subsection{Horizon, Killing vectors and Boundary Conditions}\label{Bconditions}

Considering that for black holes geometries it will be assumed that $\Sigma$ has two boundaries,
the spatial infinity $\partial\Sigma_\infty$ and the horizon $\partial\Sigma_H$, any vector fields
$\xi$ to properly define a symmetry simultaneously should preserve boundary conditions at both
surfaces. Actually the natural choice is to consider the null generator of the event horizon
\cite{Jacobson:1993xs,Wald:1993nt,Wald:1984rg}, i.e., $\xi$ is the Killing vector which is normal
to the horizon surface\cite{hawking}. Unlike the asymptotically flat case, $\xi$ can not be
normalized such that $\xi^{\mu}\xi_{\mu}=-1$ at $\partial\Sigma_{\infty}$, thus it is normalized
through the definition of the conserved charges. The vector field $\xi$ has in general the form
$\xi=\partial_{t}+\Omega^{I}\partial_{\phi^{I}}$, where $\phi^{I}$ are the independent directions
in which the angular momentum can be defined. $\Omega_{I}$ are the corresponding angular
velocities of the horizon in those directions\cite{Hawking:1998kw}.

On the other hand, the definition of boundary conditions for a theory of gravity as
Eq.(\ref{initialaction}), properly constrained to have a single cosmological constant, has been
recently discussed in \cite{Aros:1999id}. There it is introduced the \textit{ALAdS boundary
condition}, which defines that LL Lagrangian must be supplemented by the $2n$ dimensional Euler
density\footnote{Euler density is a topological invariant, therefore it does not contribute to the
field equations.},
\begin{equation}\label{EulerDensity}
  {\mathcal{E}}_{2n} = \int_{\mathcal{M}} R^{a_{1} a_{2}}\ldots
  R^{a_{2n-1} a_{2n}} \epsilon_{a_{1}\ldots a_{2n}},
\end{equation}
times a factor that depends on the particular LL Lagrangian to be considered.

This new LL Lagrangian defines a proper action principle for ALAdS spaces, since the new boundary
term $\Theta(\delta w^{ab}, w^{ab}, e^{a})$ vanishes for $\bar{R}^{cd}=R^{cd} + l^{-2} e^c e^d
\rightarrow 0$ at ${\mathbb{R}} \times \partial\Sigma_{\infty}$. This boundary condition is alike
to the electromagnetic condition, where one requires that the fields, in this case $\bar{R}^{cd}$,
to vanish at infinity \cite{Aros:1999id}.

Nonetheless, according to Ref.\cite{Wald:1993nt} to realize the black hole {\it thermodynamics} one
needs to consider the event horizon as an internal boundary of the manifold ${\mathcal{M}}$. To
take in both arguments, ALAdS condition and event horizon being a boundary, forces to set a
boundary condition at the event horizon, which does not over determine the first order field
equations. One solution to this problem corresponds to fix the spin connection $w^{ab}$ at the
event horizon. In practice this is equivalent to fix the temperature of a black hole, foreseing
that this boundary condition corresponds from a thermodynamical point of view to have been chosen
the (grand) canonical ensemble.

\section{Thermodynamics with ALAdS Condition}\label{FirstLawSection}

In general for any LL Lagrangian (\ref{initialaction}) one can demonstrate that the presymplectic
form (See Eq.(\ref{VariationofTheCharge})) is given by,
\begin{equation}\label{GravitationalSymplecticForm}
  \Xi(\delta_{1} w^{ab},-\lie_\xi w^{ab}, e^{a})= \int_{\partial\Sigma}  I_\xi \Theta(\delta_{1} w^{ab},w^{ab},e^a)
  -\delta_{1}\left(I_\xi w^{ab} \tau_{ab}\right).
\end{equation}
It is remarkable and useful that for such a general theory the presymplectic form depends only on
boundary information.

The thermodynamical relation are constructed form this last expression through the introduction of
a particular kind of variation as follows. Inside $\bar{\mathcal{F}}$ there are families of
solutions $\phi(x,\mu)$, labelled by parameters $\mu$'s. These $\mu$'s represent mass, angular
momentum, or any extensive property of the solution. Afterward, for a particular family of solution
one defines the new kind of variation,
\begin{equation}\label{massvariation}
\tilde{\delta}\phi(x, \mu)=\frac{\partial \phi(x, \mu)}{\partial \mu} \delta \mu.
\end{equation}

In expression (\ref{GravitationalSymplecticForm}) computations at infinity are considered as the
variation of Hamiltonian charges, i.e., mass, angular momentum, etc. Therefore,
\begin{equation}\label{GeneralVariationFirstOrder}
\tilde{\delta} Q_\xi = \int_{\partial\Sigma_\infty} I_\xi \Theta(\tilde{\delta}
w^{ab},w^{ab},e^a)-\tilde{\delta}\left(I_\xi w^{ab} \tau_{ab}\right).
\end{equation}

Unlike the metric formalism, subject to the Dirichlet condition, where Hamiltonian and Noether
charges are not directly related, under the ALAdS condition $\Theta(\tilde{\delta}
w^{ab},w^{ab},e^a)|_{\partial\Sigma_\infty}$ vanishes for arbitrary $\tilde{\delta} w^{ab}$.
Therefore the previous expression simply becomes
\begin{equation}\label{FirstOrderChargeVariation}
\tilde{\delta} Q_\xi = -\int_{\partial\Sigma_\infty} \tilde{\delta}\left(I_\xi w^{ab}
\tau_{ab}\right),
\end{equation}
demonstrating that the variations of the Hamiltonian and Noether charges are equivalent in this
case. This result explains the success of the Noether charge, introduced in Ref.\cite{Aros:1999id},
to reproduce the Hamiltonian charges.

It can be argued, on the other hand, that $\Xi(\tilde{\delta} w^{ab},-\lie_\xi w^{ab}, e^{a})$ must
vanishes since $\xi$ generates an isometry of the solution\cite{Lee:1990nz}. This fact was used in
Ref.\cite{Wald:1993nt} to obtain from this formalism the first law of black hole thermodynamics.
Therefore
\begin{equation}\label{boundaryFirstOrderFinalVariation}
\int_{\partial\Sigma_\infty} \tilde{\delta}\left(I_\xi w^{ab} \tau_{ab}\right) =
\int_{\partial\Sigma_H} \tilde{\delta}\left(I_\xi w^{ab} \tau_{ab}
 \right) - I_\xi \left(\tilde{\delta} w^{ab}\tau_{ab}\right).
\end{equation}

By regarding Ref.\cite{Aros:1999kt}, Eq.(\ref{FirstOrderChargeVariation}) corresponds the sum of
the variations of mass and angular momentum, therefore Eq.(\ref{boundaryFirstOrderFinalVariation})
must contain the first law of thermodynamics for any of the black hole solutions of the
gravitational theories introduced in subsection \ref{Kgravities} (See appendix B in
Ref.\cite{Crisostomo:2000bb}). Unfortunately, unlike Ref.\cite{Wald:1993nt}, in this case is not
direct to integrate the r.h.s. of Eq.(\ref{boundaryFirstOrderFinalVariation}) to obtain a formal
expression for Entropy. An alternative approach will be discussed in the next section.

To understand how Eq.(\ref{boundaryFirstOrderFinalVariation}) contains thermodynamics, it will be
analyzed for Born-Infeld action in $2n$ dimensions coupled with an electromagnetic field. The
Lagrangian of this theory is given by
\begin{equation}\label{BIaction}
{\mathbf{L}} = \hat{\kappa}\bar{R}^{a_1 a_2}\ldots\bar{R}^{a_{2n-1} a_{2n}}\epsilon_{a_1\ldots
a_{2n}} - \frac{1}{2\varepsilon \Omega_{2n-2}} F\wedge\hodge{F}
\end{equation}
where $\bar{R}^{ab}= R^{ab} + l^{-2} e^{a} e^{b}$ and
\begin{equation}\label{kappa}
  \hat{\kappa}=\frac{l^{2}}{4n(2n-2)!\Omega_{2n-2}G_{2n}}.
\end{equation}
The geometry to be considered is given by
\begin{equation}
ds^{2}=-\Delta (r)^{2}dt^{2}+\frac{dr^{2}}{\Delta (r)^{2}}+r^{2}d\Omega _{2n-2}^{2},
\end{equation}
with ${\displaystyle \Delta (r)^{2}=1+ \frac{r^{2}}{l^{2}}-\left( \frac{2MG_{2n}}{r}-
\frac{Q^2G_{2n}\varepsilon}{(2n-3)r^{2n-2}} \right)^{\frac{1}{n-1}} }$. The largest root of
$\Delta(r)$, corresponding to the outer -event- horizon, will be denoted $r_{+}$. The
electromagnetic field is given by $F=dA$, with ${\displaystyle A=\frac{\varepsilon
Q}{(2n-3)r^{2n-3}} dt}$.

Since this is not only a gravitational configuration is necessary to include into the calculations
the electromagnetic action. In order to do that it is also necessary to introduce boundary
conditions for the electromagnetic field. In the spirit of the former boundary conditions, $A$ is
fixed at the horizon $\partial\Sigma_H$  and $F$ is such that it vanishes at
$\partial\Sigma_\infty$. The vector field that defines the horizon, the Killing vector
$\xi=\partial_t$, defines the only non trivial relation. After a straightforward calculation one
obtains that additionally to the Gravitational $\Xi_{G}$ Eq.(\ref{GravitationalSymplecticForm}), it
is also presented a contribution from the electromagnetic Lagrangian, $\Xi_{E}$. Therefore, $\Xi =
\Xi_{G} + \Xi_{E}$ with $\Xi_{E}$ given by
\begin{equation}\label{BIplusElectro}
\Xi_{E} = -\frac{1}{\varepsilon\Omega_{d-2}}\int_{\partial\Sigma} f \tilde{\delta} \hodge{F} =
-\varepsilon \left.\frac{Q\tilde{\delta} Q}{r^{2n-3}}\right|^{\infty}_{r_{+}}
\end{equation}
where $f={\displaystyle \frac{\varepsilon Q}{(2n-3)r^{d-3}}}$.

Next one has to calculate the non asymptotical expression of $\tilde{\delta} w^{ab}\tau_{ab}$ and
$\tilde{\delta} I_\xi w^{ab} \tau_{ab}$ for this configuration. The results are given by
\begin{equation}\label{Result1}
\int_{\partial\Sigma_H} \tilde{\delta} \left(I_\xi w^{ab} \tau_{ab}\right)- \int_{\partial\Sigma_H}
I_\xi (\tilde{\delta} w^{ab}\tau_{ab})= \tilde{\delta} M  +\varepsilon \frac{Q\tilde{\delta}
Q}{r_+^{2n-3}}.
\end{equation}

Thus, one has that $\Xi=\Xi_{G}+\Xi_{E}={\displaystyle \left.\tilde{\delta} M
\right|^\infty_{r_{+}} }= 0$. This result confirm our previous statement that $\Xi$ must vanish in
this case. The correct way to interpret $\Xi=0$ is given by the identity which holds at the horizon
\begin{eqnarray}\label{Mr+}
\tilde{\delta} M(r_{+},Q) &=&
\left(\frac{1}{2G_{2n}}\left(1+\frac{r_{+}^2}{l^2}\right)^{n-2}\left(1+(2n-1)\frac{r_{+}^2}{l^2}\right)
-\frac{\varepsilon Q^{2}}{2r_{+}^{2n-2}} \right)\tilde{\delta} r_{+}\nonumber\\
 &+& \frac{\varepsilon Q}{(2n-2)r_+^{2n-3}} \tilde{\delta} Q,
\end{eqnarray}
where through to identify the temperature of the black hole as
\begin{eqnarray}
T &=& \frac{1}{4\pi}\left.\frac{d}{dr}\Delta(r)^2\right|_{r=r_{+}}\nonumber\\
&=& \frac{1}{4\pi(n-1)r_{+}}\left[1+(2n-1)\frac{r_{+}^{2}}{l^{2}}-\frac{G_{2n}\varepsilon
Q^2}{r_{+}^{2n-2}}\left(1+\frac{r_{+}^{2}}{l^{2}}\right)^{-(n-2)}\right]\label{Temperature},
\end{eqnarray}
yields to
\begin{equation}\label{ThermoSecondLaw}
\tilde{\delta} M \equiv \frac{\varepsilon Q}{(2n-3)r_+^{2n-3}}\tilde{\delta} Q + T\tilde{\delta}
\mathcal{S}.
\end{equation}
Here ${\mathcal{S}}$ represents the entropy of the black hole defined by
\begin{equation}\label{deltaS}
\tilde{\delta} {\mathcal{S}}=\left(1+\frac{r_{+}^{2}}{l^{2}}\right)^{(n-2)}
\frac{2\pi}{G_{2n}}(n-1)r_{+}\tilde{\delta} r_{+}.
\end{equation}
Therefore relation (\ref{ThermoSecondLaw}) resumes that $ T\tilde{\delta}
{\mathcal{S}}=\tilde{\delta} M - f_H \tilde{\delta} Q$, which represents the first law of
thermodynamics written in terms of the \textit{grand canonical ensemble}. This result was
previously  obtained in \cite{Banados:1994ur,Crisostomo:2000bb} using the standard Hamiltonian mini
super space technics.

\section{Entropy}

The relation between statistical mechanics, in the (grand)-canonical ensemble, and the path
integral states that the partition function ${\mathbf{Z}}$, for any theory of fields, can be
approximate as,
\begin{equation}\label{Z}
  -\ln({\mathbf{Z}}) =  {\mathcal{S}}-\beta E- \beta\Omega^{I} \delta Q_{I}
  \approx {\mathbf{I^{E}}}|_{\tau=\beta}.
\end{equation}
where ${\mathbf{I^{E}}}$ stands for the Euclidean version of the action, and $\Omega^{I}$ is the -
chemical -potential for the charge $Q_{I}$\cite{Hawking:1998kw}. It worth noting that Euclidian
action ${\mathbf{I^{E}}}$ has the opposite sign of the Lorentzian action ${\mathbf{I}}$, therefore
there is a global sign in the charges between Euclidean and Lorentzian frameworks.

For a gravitational LL theory in absence of matter it is satisfied that
\begin{equation}\label{start}
\delta {\mathbf{I^{E}}} = \int_{\partial {\mathcal{M}}} \delta w^{ab} \tau_{ab}.
\end{equation}
Afterward, considering that the boundary $\partial{\mathcal{M}}=\partial \Sigma \times
{\mathbb{R}}$ is defined by the time like direction times the transversal directions,
Eq.(\ref{start}) can be written as
\begin{equation}\label{FisrtVariation}
  \delta {\mathcal{S}}-\beta \delta E
  - \beta\Omega^{I} \delta J_{I} = \beta \int_{\partial \Sigma} I_{\chi} \left( \delta w^{ab}
  \tau_{ab}\right),
\end{equation}
where $\chi$ is the Euclidian version of $\xi$, therefore
$\chi=\partial_{\tau}+\Omega^{I}\partial_{\varphi^{I}}$\cite{Hawking:1998kw}, being
$(\tau,\varphi^{I})$ the wick rotated version of $(t,\phi^{I})$.

By recalling that under ALAdS condition the integral at $\partial\Sigma_{\infty}$
vanishes\footnote{Even though in the Euclidian framework the horizon is not a surface but a point
expressions are formally correct.},
\begin{eqnarray}
   \delta {\mathcal{S}}-\beta \delta E - \beta\Omega^{I} \delta J_{I} &=&  -\beta \int_{\partial\Sigma_{H}} I_{\chi} \left( \delta w^{ab} \tau_{ab}\right)\\
   &=& -\beta \int_{\partial\Sigma_{H}} \left[ I_{\chi} \left( \delta w^{ab} \tau_{ab}\right)-
    \delta\left(  I_{\chi}w^{ab} \tau_{ab}\right)\right]+ \delta\left(  I_{\chi}w^{ab}
    \tau_{ab}\right).
\end{eqnarray}
However, at the r.h.s. it can be used the Euclidian version of
Eq.(\ref{FirstOrderChargeVariation}),
\begin{equation}
 \int_{\partial\Sigma_{H}} I_{\chi} \left( \delta w^{ab} \tau_{ab}\right)-
    \delta\left(  I_{\chi}w^{ab} \tau_{ab}\right) =  -\int_{\partial\Sigma_{\infty}}\delta\left(  I_{\chi}w^{ab}
    \tau_{ab}\right) = \delta E + \Omega^{I} \delta J_{I}.
\end{equation}
Thus one has that
\begin{equation}\label{Entropy}
\delta {\mathcal{S}}-\beta \delta E- \beta\Omega^{I} \delta J_{I} = \beta \int_{\partial\Sigma_{H}}
\delta\left(I_{\chi} w^{ab} \tau_{ab}\right) - \beta \delta E - \beta\Omega^{I} \delta J_{I},
\end{equation}
which yields to
\begin{equation}\label{S}
 {\mathcal{S}} = {\mathcal{S}}_{0}+ \beta \int_{\partial\Sigma_{H}} I_{\chi} w^{ab} \tau_{ab},
\end{equation}
where ${\mathcal{S}}_{0}$ is an integral constant. Finally one can check that the internal product
$(I_{\chi} w^{ab})$ is invariant under the Wick rotation, therefore Eq.(\ref{S}) is still valid in
the Lorentzian framework.

It can be confirmed that Eq.(\ref{S}) reproduces the standard results, modulo an integral
constant. For instance, it reproduces the result of Eq.(\ref{deltaS}), and in four dimensions the
area law is obtained by choosing ${\mathcal{S}}_{0}=-(\pi l^{2})/G$ for Schwarzschild-AdS and
${\mathcal{S}}_{0}=-(\pi l^{2})/(2G)$ for Kerr-AdS.

It is worth noting that in Ref.\cite{Wald:1993nt} ( Eq.(26)) is given a similar expression for the
entropy for an asymptotically flat space. One can compare both expressions by noting that
expression under integration in Eq.(\ref{S}) in principle can be connected with its analogous in
Ref.\cite{Wald:1993nt}. On the other hand ${\mathcal{S}}_{0}$ acts as a pseudo regulator parameter
that allows to take the \textit{flat limit} ($l\rightarrow\infty$) in the final expression for the
entropy. The presence of pseudo regulator parameters in the theory has been already noted in
Ref.\cite{Aros:1999id} where the Euler density was interpreted in this way.

\section{Conserved Charge Algebras}
In this section is introduced the analysis of the algebra of asymptotical charges by omitting the
existence of an internal boundary. By restricting the presymplectic form to the space of classical
solutions\cite{Lee:1990nz,Carlip:1999cy} it can be interpreted as the Dirac bracket of two
conserved charges. This is equivalent to the Hamiltonian procedure where after gauge fixing the
Poisson bracket of two Generators, supplemented by the charges that make them differentiable,
corresponds to the Dirac bracket of the charges -evaluated on the solutions-
\cite{Regge:1974zd,Benguria:1977in}. Therefore,
\begin{equation}\label{algebra}
  [Q_1,Q_2] = \int_{\Sigma_{\infty}} \hat{\delta}_2 \Theta(\hat{\delta}_1)-\hat{\delta}_1
  \Theta(\hat{\delta}_2).
\end{equation}

This \emph{bracket} will define an extension of the algebra of symmetries.

\subsection{Lorentz Algebra}
A Lorentz transformation is given in terms of the fields as
\begin{equation}\label{LorentzTransformation}
  \hat{\delta}_{\lambda} e^a = \lambda^a_b e^b\mbox{       and          }\hat{\delta}_{\lambda} w^{ab}J_{ab} =
 -D(\lambda^{ab}J_{ab}),
\end{equation}
which through the Noether method defines the conserved current
\begin{equation}\label{LocalLorentzCurrnet}
  \hodge{{\bf J}_\lambda} = \Theta(\hat{\delta}_{\lambda} w^{ab}, w^{ab},e^a) = -D(\lambda^{ab}) \tau_{ab}
\end{equation}
and subsequently the Noether charge
\begin{equation}\label{LocalLorentzCharge}
  Q^{N}_{\lambda} = -\int_{\partial \Sigma_{\infty}} \lambda^{ab} \tau_{ab}.
\end{equation}

On the other hand, one knows that the variation of the Hamiltonian charge is given by,
\[
\tilde{\delta} Q^{H}_{\lambda} = \Xi(\tilde{\delta} w^{ab}, \hat{\delta}_{\lambda} w^{ab}) =
-\int_{ \Sigma}
 \tilde{\delta} (D(\lambda^{ab})\tau_{ab})+\hat{\delta}_{\lambda} (\tilde{\delta} w^{ab} \tau_{ab})  ,
\]
however the last term of the r.h.s. vanishes since $\tilde{\delta} w^{ab} \tau_{ab}$ is a scalar
for the Lorentz group. Using this result is straightforward to identify Noether and Hamiltonian
charges.

In order to determinate the algebra of these charges, one calculates expression (\ref{algebra}) for
two independent Lorentz transformations, $\hat{\delta}_{\lambda_1}$ and $\hat{\delta}_{\lambda_2}$.
Therefore,
\begin{equation}\label{LorentzAlgebra} [Q_{\lambda_1},Q_{\lambda_2}] = \int_{\partial
\Sigma_{\infty}} [\lambda_1,\lambda_2]^{ab} \tau_{ab}=Q_{[\lambda_1,\lambda_2]},
\end{equation}
where $[\lambda_1,\lambda_2]$ stands for the bracket between the parameters of the transformations
in the Lorentz algebra. This result shows that the algebra of charges reproduces the algebra of
local Lorentz transformations.

\subsection{Diffeomorphism Algebra}
The next step is to analyze the algebra of charges associated with the invariance under
diffeomorphisms. In general that can be done by simply restricting the analysis to the asymptotic
behavior of the space, thus, it is only necessary to consider $\xi$ as an asymptotical Killing
vector. During section \ref{FirstLawSection} (See Eq.(\ref{FirstOrderChargeVariation}) ) was
exhibited that under the ALAdS condition Hamiltonian and Noether charges agree. Therefore the
algebra of charges is given by \footnote{ In this calculation was useful the identity
$I_{\eta}\lie_{\zeta}- I_{\zeta}\lie_{\eta} = I_{[\eta,\zeta]}$}
\begin{eqnarray}
[Q_{\zeta},Q_{\xi}] &=& \int_{\Sigma_{\infty}} -\lie_{\zeta} \Theta(-\lie_{\xi})+ \lie_{\xi}
\Theta(-\lie_{\zeta})\nonumber\\
&=& Q_{[\zeta,\xi]} + \int_{\partial \Sigma_{\infty}} I_\zeta w^{ab} \lie_{\xi} \tau_{ab} -  I_\xi
w^{ab} \lie_{\zeta} \tau_{ab},
\end{eqnarray}
where one recognizes an additional term,
\begin{equation}\label{FinalDiffeomorphismCommutationRelation}
K(-\lie_\zeta,-\lie_\xi) = \int_{\partial \Sigma_{\infty}} I_\zeta w^{ab} \lie_{\xi} \tau_{ab} -
I_\xi w^{ab} \lie_{\zeta} \tau_{ab}.
\end{equation}
It is matter of further investigation to establish whether this former expression is an extension
of the algebra of diffeomorphisms and what is its classification. To accomplish that previously it
has to be understood which are the families of diffeomorphisms that are compatible with the ALAdS
condition. Nonetheless, it can be analyzed as a very particular case a locally AdS
space\footnote{Note that a locally AdS space has constant curvature everywhere. Nonetheless, there
are several non-trivial spaces which are locally AdS in four dimensions. See
\cite{Aminneborg:1997pz,Brill:1997mf,Aros:2000ij} and references therein.}. In this particular
geometry one can expect that $K(-\lie_\zeta,-\lie_\xi)=0$. In four dimensions, for improved  EH
action (See \cite{Aros:1999id} for details) one obtains that
\begin{equation}\label{EHCentralTerm}
K(\zeta,\xi) =  \frac{l^2}{64\pi G}\int_{\partial \Sigma_{\infty}} I_\zeta w^{ab} \lie_{\xi}
\bar{R}^{cd} \epsilon_{abcd} - I_\xi w^{ab} \lie_{\zeta}\bar{R}^{cd} \epsilon_{abcd}
\end{equation}

However, given that $\bar{R}^{cd}=R^{cd} + l^{-2} e^c e^d =0$, one can check that the former
expression vanishes. Therefore, one can conclude that for any locally AdS manifold in four
dimensions, the algebra of diffeomorphism does not have an extension.

This result can be generalized to higher (even) dimensions and to any of the LL Lagrangians
discussed. This can be done by noting that under the ALAdS condition $\tau_{ab}$ is proportional
to a power of $\bar{R}^{ab}$ and thus it vanishes everywhere evaluate on a locally AdS space.
Thus, that determinates that $K(-\lie_\zeta,-\lie_\xi)$ (See
Eq.(\ref{FinalDiffeomorphismCommutationRelation})) vanish for any {\bf even dimensional locally
AdS space}.

\section{Conclusions and Open Questions}

In this work was performed a first approach to the analysis of the phase space of the first order
gravitational theories subject to the ALAdS condition. It was demonstrated that this boundary
condition ensures that black hole thermodynamics holds for ALAdS spaces, even in presence of
matter. It is clarified the reason for Hamiltonian and Noether methods match in this case.

It is remarkable that after only a fast analysis it is possible to gain information about the
nature of the conserved charges algebras. One may expect that, as it can be done for the metric
formalism in $2+1$ dimensions \cite{Carlip:1999cy,Brown:1986nw}, degenerate states can be analyzed
by examining classes of the diffeomorphisms that preserve the boundary conditions. \vspace{0.2in}

I am grateful to J. Cris\'{o}stomo, C. Mart\'{\i }nez, O. Mi\v{s}kovi\'{c}, M.
Plyushchay, J. Saavedra, C. Teitelboim, and  R. Troncoso. by many enlightening discussions. In
particular I would like to thank to M. Contreras, R. Olea and J. Zanelli for helping remarks. Also
I would like to thank the "Service de physique th\'eorique et math\'ematique de l'Universit\'e
Libre de Bruxelles" for kind hospitality.

This work was supported in part through grants 3990009 from FONDECYT, and grant DI 51-A/99 (UNAB).
The institutional support of a group of Chilean companies (CODELCO, Dimacofi, Empresas CMPC, MASISA
S.A. and Telef\'onica del Sur) is also acknowledged. CECS is a Millennium Science Institute.


\end{document}